\documentclass[preprint]{elsarticle} 

\usepackage[utf8]{inputenc}
\usepackage[english]{babel}
\usepackage[T1]{fontenc}
\usepackage{amsmath}
\usepackage{hyperref}
\usepackage{amssymb}
\usepackage{graphicx}
\usepackage{braket}
\usepackage{algorithm}
\usepackage{algpseudocode}
\usepackage{float} 
\usepackage[dvipsnames]{xcolor}
\usepackage{amsthm}
\usepackage{qcircuit}
\usepackage{hyperref}
\usepackage{tikz}
\usetikzlibrary{positioning, shapes, calc}

\usepackage[capitalise]{cleveref}
\usepackage{parskip}
\usepackage{comment}
\usepackage{multirow}
\usepackage{wrapfig}

\newtheorem{theorem}{Theorem}
\newtheorem{lemma}{Lemma}

\newtheorem{corollary}{Corollary}

\newcommand{\titlebox}[2]{
\begin{tikzpicture}
\node[draw, rectangle, thick,inner sep=3mm, align=left, text width=\textwidth-6mm] (titlebox) {#2};
\node[fill=white] (Title) at ($(titlebox.north) + (-0.25\textwidth, 0)$) {\bfseries #1};
\end{tikzpicture}
}

\newcommand{\arrow}[1]{\xrightarrow[]{#1}}

\newcommand{\SGate}{\text{S}}
\newcommand{\HGate}{\text{H}}
\newcommand{\CXGate}{\text{CX}}
\newcommand{\XGate}{\text{X}}

\newcommand{\IGate}{\text{I}}

\newcommand{\TGate}{\text{T}}

\newcommand{\TOF}{\text{TOF}}

\newcommand{\ComplexityClass}[1]{\mathbf{#1}}
\newcommand{\PTIME}{\ComplexityClass{P}}
\newcommand{\NP}{\ComplexityClass{NP}}

\newcommand{\NQP}{\ComplexityClass{NQP}}

\newcommand{\CEQP}{\ComplexityClass{C}_{=}\PTIME}

\newcommand{\PH}{\ComplexityClass{PH}}

\newcommand{\CO}[1]{\text{co-}#1}

\newcommand{\gateset}{\mathcal{A}}
\newcommand{\gatesetB}{\mathcal{B}}

\newcommand{\COUNTSAT}[1]{|\{x:#1(x)\}|}
\newcommand{\DeutschJozsa}[1]{\text{DJ}_{#1}}

\newcommand{\DJGADGET}{Q}

\newcommand{\ISIDENTITYPROMISE}[1]{\textsc{Promise\text{-}IsIdentity}(#1)}
\newcommand{\ISIDENTITYPROMISEBOLD}[1]{\textbf{\textsc{Promise\text{-}IsIdentity}}\textbf{(}#1\textbf{)}}

\newcommand{\OPTIMIZATIONPROBLEM}[1]{\textsc{Exact\text{-}Optimization}(#1)}

\newcommand{\ISBALANCED}{\textsc{IsBalanced}}

\newcommand{\Paragraph}[1]{\smallskip\noindent{\bf #1}}
\newcommand{\SubParagraph}[1]{\smallskip\noindent{\em #1}}

\title{Exact Quantum Circuit Optimization is co-NQP-hard}

\author[1]{Adam Husted Kjelstrøm\corref{cor1}}
\ead{husted@cs.au.dk}

\author[1]{Andreas Pavlogiannis}
\ead{pavlogiannis@cs.au.dk}

\author[1]{Jaco van de Pol}
\ead{jaco@cs.au.dk}

\cortext[cor1]{Corresponding author}
\affiliation[1]{organization={Aarhus University, Dept.\ of Computer Science},
addressline={Åbogade 34},
postcode={8200},
city={Aarhus N},
country={Denmark}}

\begin{document}

\begin{abstract} 
    As quantum computing resources remain scarce and error rates high, minimizing the resource consumption of quantum circuits is essential for achieving practical quantum advantage.
    Here we consider the natural problem of, given a circuit $C$, computing a circuit $C'$ that behaves equivalently on a desired subspace, and that minimizes a quantum resource type, expressed as the count or depth of
    (i)~all gates, or
    (ii)~non-Clifford gates, or
    (iii)~superposition gates, or
    (iv)~entanglement gates.
    We show that, when $C$ is expressed over any gate-set that can implement the $\HGate$ and $\TOF$ gates exactly, each of the above optimization problems is hard for $\CO{\NQP}$, and hence outside the Polynomial Hierarchy, unless the Polynomial Hierarchy collapses. 
    This complements a recent result of Van de Wetering and Amy (arXiv, 2023) which establishes an $\NP$-hardness lower bound when equivalence is over the full state space,
    and tightens the gap to the corresponding $\NP^{\NQP}$ upper bound known for cases (i)-(iii) over Clifford+$\TGate$ and (i)-(iv) over $\HGate$+$\TOF$ circuits. 
    Our result also generalizes a result by Tanaka (Int. J. Quantum Inf., 2010) to other gate-sets and quantum resources.
\end{abstract}


\maketitle

\section{Introduction}

Despite rapid progress in quantum hardware, limited resources and high error rates remain fundamental bottlenecks, making the optimization of quantum circuits a key challenge on the path to quantum advantage~\cite{Karuppasamy2025QuantumCircuitOptimization}.
Quantum computations rely on multiple types of resources, and depending on the underlying hardware, different resources (e.g., the total circuit size, or the number of gates of a particular type) can become the primary bottleneck by introducing noise or increasing execution time~\cite{Muroya2025HardwareOptimal}.

At a high level, a circuit optimization problem is defined with respect to an objective function over circuits.
Then, given a circuit $C_\gateset$ over some fixed gate-set $\gateset$, the task is to find a circuit $C_\gatesetB$ over some gate-set $\gatesetB$ (possibly different from $\gateset$) such that $C_\gateset$ and $C_\gatesetB$ are equivalent, and the objective function is minimized.
Here we focus on exact equivalence as considered in~\cite{Tanaka2010ExactNonIdentityCheckNQPComplete,Wetering2023OptimisingQuantumCircuitsHard}, where $C_\gateset$ and $C_\gatesetB$ behave equally on a specified subspace of input states.
Although quantum computation is often approximate~\cite{Dawson2005solovay_kitaev_thrm,Nielsen_Chuang2010}, focusing on exact equivalence remains important:~it is the direct quantum analogue of classical computing, and it is useful for canonicalizing circuits, i.e., building a database of the most efficient circuits~
\cite{Bravyi2022_6_qubit_optimal_Clifford_circuits}, making fast peephole optimizations compositional (i.e., without the need to keep track of how approximation errors may accumulate).

Exact quantum circuit optimization has been studied with respect to the Clifford fragment~\cite{shaik2025cnotoptimalcliffordsynthesissat,Schneider2023}, a non-universal fragment for quantum computing~\cite{Aaronson04sim_of_stabilizer_circuits}.
To capture the full computational power of quantum computing, the circuit optimization problem has also been studied with respect to (approximately) universal gate-sets beyond the Clifford fragment.
Tanaka~\cite{Tanaka2010ExactNonIdentityCheckNQPComplete} showed that there are fixed gate-sets for which even deciding whether a circuit behaves as the identity on a subspace (i.e., with respect to the behavior on this subspace, it can be implemented with $0$ gates) is $\CO{\NQP}$-hard under Karp reductions.
More recently, Van de Wetering and Amy~\cite{Wetering2023OptimisingQuantumCircuitsHard} studied the optimization problem for fixed gate-sets and with respect to minimizing either the circuit size/depth or the count/depth of some particular gate $G$ (from now on referred to as $G$-count/depth), over the full input space.
In particular, the optimization problem was shown to be $\NP$-hard, under $\PTIME$-time Turing reductions, over the Clifford+$\TGate$ and $\HGate$+$\TOF$ gate-sets, 
and where minimization is with respect to the circuit size/depth or the $\HGate$/$\CXGate$/$\TGate$-count/depth for the former, and the $\HGate$/$\TOF$-count/depth for the latter.
The same work also showed that these problems are in $\NP^{\NQP}$ (for both the full input space and strict subspaces), except for the case of $\CXGate$ gates, for which an upper bound is not known.
As $\NQP$ is believed to be outside the Polynomial Hierarchy $\PH$~\cite{fenner1998determiningacceptancepossibilityquantum}, there is a large gap between the $\NP^{\NQP}$ upper bound and the $\NP$-hardness lower bound.
To our knowledge, the complexity of minimization with respect to all resources except total count for subspace equivalence is unknown.
How difficult is exact quantum circuit optimization?

In this work, we consider the optimization problem in the following general form.
Fix gate-sets $\gateset$ and $\gatesetB$. 
Then given a circuit $C_\gateset$ over $\gateset$, the task is to construct a circuit $C_\gatesetB$ over $\gatesetB$ that agrees with $C_\gateset$ on a desired subspace while minimizing one of the following resources:
(i)~the number of all gates, or 
(ii)~the number of non-Clifford gates (e.g., $\TGate$ gates), or
(iii)~the number of gates that generate superposition (e.g., $\HGate$ gates), or
(iv)~the number of gates that generate entanglement (e.g., $\CXGate$ gates).
While case (i) deals with the standard objective of minimizing circuit size, cases (ii), (iii) and (iv) deal with the minimization of various resources that are needed for universal quantum computation~\cite{Aaronson04sim_of_stabilizer_circuits}, but are more general than the count/depth of specific gates as studied in~\cite{Wetering2023OptimisingQuantumCircuitsHard}.
We show that, in all cases, the optimization problem is $\CO{\NQP}$-hard, and thus likely outside $\PH$.
This hardness holds for arbitrary $\gatesetB$ which need not even implement the gates of $\gateset$.
Towards this, we prove $\CO{\NQP}$-hardness for a conceptually simpler, promise problem, defined in~\cref{sec:definitions}, that unifies the hardness of the above optimization tasks.

The remainder of the paper is structured as follows.
In \cref{sec:definitions}, we outline standard concepts and notation in quantum circuits, characterize the gates a circuit must contain in order to produce each of the resources we consider, and define our intermediate promise problem.
In \cref{sec:DJ_gadget}, we introduce a Deutsch-Jozsa gadget that serves as the basis of our reduction later, and establish some key properties of its unitary.
Finally, in \cref{sec:hardness}, we prove the $\CO{\NQP}$-hardness of our promise problem, and prove its consequences on the complexity of exact circuit optimization.

\section{Definitions}\label{sec:definitions}

Throughout the paper, we use $a,b,c \in \{0,1\}$ to denote single (classical) bits,
and $x,y,z,w \in\{0,1\}^n$ to denote bit-strings of length $n$, where $n$ is either explicit or clear from the context, and $x_j$ gives the $j$'th bit of $x$.
We denote by $x\cdot y$ the bitwise inner product of $x$ and $y$, modulo two, and by $a\oplus b$ the sum of $a$ and $b$, modulo two.

\Paragraph{Quantum states and gates.}
We assume familiarity with the basics of quantum computing, and refer to~\cite{Nielsen_Chuang2010,Qiu2025} for an introduction.
An $n$-qubit quantum state is a unit vector $\ket{\psi} = \sum_{x\in\{0,1\}^n} \psi_x \ket{x}$ in $\mathbb{C}^{2^n}$, where $\{\ket{x}\}_{x\in\{0,1\}^n}$ is the computational basis, and measuring $\ket{\psi}$ yields outcome $x$ with probability $P(x)=|\psi_x|^2$.
For a quantum state $\sum_{a,x} \psi_{a,x} \ket{a,x}$ over qubit $a$ and qubits $x$, $P(a = 1|x)$, the probability of qubit $a$ measuring to $1$ given outcomes on qubits $x$, is $|\psi_{1,x}|^2/(|\psi_{0,x}|^2+|\psi_{1,x}|^2)$, which is defined only for outcomes $x$ of non-zero probability, i.e. when $|\psi_{0,x}|^2+|\psi_{1,x}|^2>0$. 
We write $\ket{\psi_1}\ket{\psi_2}$ for the (tensor) product of states $\ket{\psi_1}$ and $\ket{\psi_2}$. 
A quantum gate $G$ on $r$ qubits represents a $(2^r \times 2^r)$-dimensional unitary matrix $U_G$, which, when $G$ is applied to a subset of the qubits of a larger state, is implicitly tensored with the identity $\IGate$ on the remaining qubits.
We write $\ket{\psi}\arrow{G}\ket{\psi'}$ to mean that $U_G \ket{\psi}=\ket{\psi'}$, write $G^{\otimes k}$ for the $k$-fold tensor product of $G$ with itself, and, by linearity, specify gates by their action on basis states.

\Paragraph{Circuits.}
A circuit $C$ is a sequence of gates $G_1 G_2\dots G_m$, implementing the unitary $U_C = U_{G_m}\dots U_{G_2} U_{G_1}$. 
Let $S$ be a set of gates. 
The $S$-count of $C$ is $|\{i \in \{1,\dots,m\} \colon G_i \in S\}|$.
We call a sequence of indices $j_1 < j_2 <\dots < j_\ell$ from $\{1,\dots,m\}$ a chain of $C$, if for every $k<\ell$, the gates $G_{j_k}$ and $G_{j_{k+1}}$ act on at least one common qubit.
Note that a chain need not consist of consecutive indices: it is an arbitrary ordered selection of indices, so that $G_{j_1} \dots G_{j_\ell}$ is a (generally non-contiguous) subsequence of $C$ in which each gate is applied to a set of qubits which overlaps with the next. 
The $S$-depth of $C$ is the maximum of $|\{k\colon G_{j_k}\in S\}|$ over all chains $j_1 < \dots < j_\ell$ of $C$.
Taking $S=\{G_1,\dots,G_m\}$ recovers the usual notion of circuit depth.

Given a quantum circuit $G$, we define $\Lambda^k G$ as the $k$-controlled version of $G$, computing 
\[
\Lambda^k G \colon \ket{x}\ket{y} \mapsto \begin{cases}
    \ket{x}\ket{y} & \text{if $x\neq 1^k$} \\
    \ket{x}(U_G\ket{y}) & \text{if $x=1^k$.}
\end{cases}
\]
In words, $\Lambda^k G$ applies $G$ if and only if the control bits equal $\ket{1^k}$.

\Paragraph{Clifford gates.}
The Clifford gate-set consists of the Phase ($\SGate$), Hadamard ($\HGate$), and CNOT ($\CXGate$) gates, and constitutes an important family of quantum gates, each acting on the 1- resp. 2-qubit basis states as follows. 
\begin{align*}
    \SGate\colon\ket{a} \mapsto i^{a} \ket{a} \qquad
    \HGate\colon \ket{a} \mapsto \frac{1}{\sqrt{2}} \sum_{b=0}^1 (-1)^{ab}\ket{b} \qquad
    \CXGate\colon\ket{a,b} \mapsto \ket{a,a\oplus b}
\end{align*}
Here $i$ satisfies $i^2=-1$.
A stabilizer state is any state obtainable from a basis state by a Clifford circuit~\cite{Aaronson04sim_of_stabilizer_circuits}.
In the following, we discuss different quantum computing resources.

\Paragraph{Non-Clifford gates.}
Access to non-Clifford resources is necessary for quantum universality~\cite{Aaronson04sim_of_stabilizer_circuits}.
It is well-known that extending the Clifford gates with any gate not expressible with Clifford gates yields an (approximately) universal gate-set for quantum computing~\cite{aharonov2003simple,Dawson2005solovay_kitaev_thrm}.
Two commonly considered non-Clifford gates are the $\TGate$ and $\TOF$ gates, each acting on the 1- resp. 3-qubit basis states as follows.
\begin{align*}
    \TGate\colon\ket{a}\mapsto \sqrt{i}^a \ket{a} \qquad \qquad \TOF\colon\ket{a,b,c} \mapsto \ket{a,b,c \oplus a\wedge b}
\end{align*}

In later proofs, we use the fact that a Clifford circuit acting on a basis state produces only outcome probabilities in $\{0,1/2,1\}$~\cite{dehaene2003clifford}; we make this precise in~\cref{lemma:state_obtained_by_circuit_gives_property_of_circuit}~(iii) below. 
Accordingly, we say that a state $\ket{\psi}$ has a biased outcome probability, if there are a qubit $a$ and qubits $x$ over $\ket{\psi}$ (possibly none) with $P(a=1|x)\notin \{0,1/2,1\}$. 
An example is the single-qubit state $\tfrac{\sqrt{3}}{2}\ket{0} + \tfrac{1}{2}\ket{1}$, for which $P(a=1) = 1/4$.

\Paragraph{Entanglement gates.}
An entangled state is any state $\ket{\psi}$ which cannot be written as a product of single-qubit states, i.e., for all single-qubit states $\ket{\psi_1}$, $\dots$, $\ket{\psi_n}$, we have $\ket{\psi} \neq \ket{\psi_1}\dots\ket{\psi_n}$.
We say that a gate $G$ generates entanglement, if there exists a product state $\ket{\psi_1}\ket{\psi_2}$ such that $G$ maps $\ket{\psi_1}\ket{\psi_2}$ to an entangled state.
An example is the $\CXGate$ gate, as $\CXGate \colon \frac{1}{\sqrt{2}}(\ket{0}+\ket{1})\ket{0} \mapsto \frac{1}{\sqrt{2}}(\ket{00}+\ket{11})$. $\TOF$ is also entangling.
In contrast, $\HGate$ and $\TGate$ are not entanglement gates, as they only affect one qubit.
We say that a state $\ket{\psi}$ exhibits correlation if there are a qubit $a$ and qubits $x$ over $\ket{\psi}$ as well as bit-strings $x_1$ and $x_2$ such that 
$$P(a=1|x=x_1)\neq P(a=1|x=x_2) \ .$$ 
An example is the state $\tfrac{1}{\sqrt{2}} (\ket{00}+\ket{11})$ over $a$ and $b$, for which $P(a=1|b=1)=1$ while $P(a=1|b=0) = 0$. 
In later proofs we use the fact that a circuit which produces correlation from an uncorrelated input must contain an entangling gate; we make this precise in~\cref{lemma:state_obtained_by_circuit_gives_property_of_circuit}~(ii) below.

\Paragraph{Superposition gates.}
Superposition is a fundamental resource in quantum computing.
A superposition state is any state $\sum_{y} \psi_y \ket{y}$ with $|\{y\colon \psi_y \neq 0\}|\geq 2$, that is, it is a weighted sum of two or more basis states.
We say that a gate $G$ generates superposition, if there exists a basis state $\ket{x}$ such that $G$ maps $\ket{x}$ to a superposition state.
An example is the $\HGate$ gate, as $\HGate\colon \ket{0}\mapsto \frac{1}{\sqrt{2}}(\ket{0}+\ket{1})$.
In contrast, $\CXGate$, $\TGate$ and $\TOF$ are not superposition gates, as they map between basis states.

The definitions above define a resource gate (non-Clifford, entanglement, or superposition gate) from the way it acts on basis states. 
As the following lemma states, the opposite relation is also true: the only way to generate each of the state types is by using the corresponding resource gate.
The statement is folklore; we include a proof for completeness.

\smallskip
\begin{lemma}
    \label{lemma:state_obtained_by_circuit_gives_property_of_circuit}
    Let $D$ be a circuit acting on a basis state $\ket{x}$, and let $\ket{\chi}=U_D\ket{x}$.
    (i)~If $\ket{\chi}$ is a superposition state, then $D$ contains at least one superposition gate.
    (ii)~If $\ket{\chi}$ exhibits correlation, then $D$ contains at least one entanglement gate.
    (iii)~If $\ket{\chi}$ has a biased outcome probability, then $D$ contains at least one non-Clifford gate.
\end{lemma}

\begin{proof}
    (i)~If no gate of $D$ generates superposition, then by definition each gate maps every basis state to a basis state (possibly with a scalar). 
    By induction on the length of $D$, $U_D \ket{x}$ is a basis state (possibly with a scalar), which contradicts the assumption that $U_D\ket{x}$ is a superposition state. 
    (ii)~If no gate of $D$ generates entanglement, then each gate maps product states to product states, and since $\ket{x}$ is a product of single-qubit states, so is $\ket{\chi}$. 
    In a product state the outcomes of distinct qubits are independent, so for every qubit $a$ and every subset of qubits $x$ excluding $a$, $P(a=1|x=x_1)=P(a=1)=P(a=1|x=x_2)$ for all $x_1$ and $x_2$, so $\ket{\chi}$ exhibits no correlation. 
    (iii)~If $D$ consists only of Clifford gates, then $\ket{\chi}$ is a stabilizer state.
    By definition, $P(a=1|x)$ on $\ket{\chi}$ equals $P(a=1)$ on the state $\ket{\chi_x}$, obtained from $\ket{\chi}$ by applying measurements in the computational basis according to $x$.
    Since stabilizer states are closed under measurements in the computational basis~\cite{Aaronson04sim_of_stabilizer_circuits}, $\ket{\chi_x}$ is a stabilizer state, so $P(a=1)\in \{0,1/2,1\}$. 
    Then $P(a=1|x)\in \{0,1/2,1\}$ and $\ket{\chi}$ has no biased outcome probability.
\end{proof}

\Paragraph{Boolean predicates.}
Let $\varphi\colon \{0,1\}^n \arrow{} \{0,1\}$ be a Boolean predicate.
Let $\COUNTSAT{\varphi}$ be the number of satisfying assignments to $\varphi$.
We say that $\varphi$ is balanced, if $\COUNTSAT{\varphi} = \COUNTSAT{\neg \varphi} = 2^{n-1}$, meaning it has the same number of satisfying and unsatisfying assignments.
A Boolean formula is a Boolean predicate inductively defined over conjunctions, disjunctions, and negations.

\Paragraph{Complexity classes.}
For a complexity class $\ComplexityClass{X}$, we write $\NP^\ComplexityClass{X}$ for the class of problems decidable by a polynomial-time nondeterministic Turing machine with oracle access to a problem in $\ComplexityClass{X}$. 
The polynomial hierarchy is $\PH = \bigcup_{k \geq 0} \ComplexityClass{\Sigma}_k^p$, where $\ComplexityClass{\Sigma}_0^p = \PTIME$ and $\ComplexityClass{\Sigma}_{k+1}^p = \NP^{\ComplexityClass{\Sigma}_{k}^p}$; thus $\ComplexityClass{\Sigma}_1^p = \NP$.
We say that $\PH$ collapses if $\PH = \ComplexityClass{\Sigma}_k^p$ for some fixed $k$, which is widely believed not to be the case. 
For a class $\ComplexityClass{X}$, $\CO{\ComplexityClass{X}}$ denotes the class of complements of problems in 
$\ComplexityClass{X}$; in particular $\CO{\NQP}$ consists of the problems whose complement is decidable by a polynomial-time quantum Turing machine with acceptance defined as reaching an accepting state with non-zero amplitude~\cite{adleman1997quantum}.
It is known that $\CO{\NQP}=\CEQP$~\cite{Yamakami1999nqp}, where $\CEQP$ is the class of problems decidable by a polynomial-time non-deterministic Turing machine such that the number of accepting paths exactly equals the number of rejecting paths, if and only if the answer is `yes'~\cite{wagner1986complexity}.
It is known that $\CEQP \not\subseteq \PH$ unless $\PH$ collapses~\cite{TodaOgiwara1992}.
A canonical complete problem, under Karp reductions, for $\CEQP$ is $\ISBALANCED$:~given a Boolean formula $\varphi\colon \{0,1\}^n \arrow{}\{0,1\}$, accept if and only if $\COUNTSAT{\varphi}=\COUNTSAT{\neg \varphi}=2^{n-1}$~\cite{wagner1986complexity}.

\Paragraph{The problem $\ISIDENTITYPROMISEBOLD{\gateset}$.}
Our hardness on exact circuit optimization follows as a corollary of the hardness of the following promise problem, which we introduce here.

\titlebox{$\ISIDENTITYPROMISEBOLD{\gateset}$}{
\emph{Input:} An $n$-qubit circuit $C$ over gate-set $\gateset$, and a bit-string $x\in \{0,1\}^n$. \\
\emph{Promise:} Either $U_C\ket{x} = \ket{x}$ or $U_C\ket{x}$ is a superposition state that exhibits correlation and has a biased outcome probability.\\
\emph{Question:} Is it the case that $U_C\ket{x}=\ket{x}$?
}
Note that $\ket{x}$ is a basis state, so it is not a superposition state, nor does it exhibit correlation or a biased outcome probability, meaning the promise indeed separates the two cases.
The condition $U_C\ket{x}=\ket{x}$ can be understood as requiring that $C$ behaves as the identity on the subspace spanned by $\ket{x}$, hence the name.

\section{A Deutsch-Jozsa-Based Gadget}\label{sec:DJ_gadget}

Towards our hardness proof of $\ISIDENTITYPROMISE{\gateset}$, we introduce a circuit $\DeutschJozsa{\varphi}$, which implements the Deutsch-Jozsa algorithm~\cite{DeutschJozsa1992RapidSolutionProblemsQuantumComputation}.
The algorithm takes as input a Boolean formula $\varphi$, and under the promise that $\varphi$ is either balanced or constant (meaning all outputs are equal), the algorithm efficiently decides which of these cases applies.
The $\DeutschJozsa{\varphi}$ circuit is shown in \cref{fig:deutsch_jozsa_alg}.

\begin{figure}[H]
    \centering
\begin{center}

\begin{equation*}
\Qcircuit @C=1em @R=1.6em {
   \lstick{\ket{x}} & {/} \qw & \gate{\HGate^{\otimes n}} & \gate{U^{\pm}_\varphi} & \gate{\HGate^{\otimes n}} & \qw \\
}
\end{equation*}

\end{center}
    \caption{A circuit implementing the Deutsch-Jozsa algorithm~\cite{DeutschJozsa1992RapidSolutionProblemsQuantumComputation}.}
    \label{fig:deutsch_jozsa_alg}
\end{figure}

Here $U^\pm_\varphi$ is a standard means of giving quantum access to $\varphi$ by computing $U^\pm_\varphi\colon \ket{x}\mapsto (-1)^{\varphi(x)} \ket{x}$.
$U^\pm_\varphi$ can be implemented exactly over $\HGate$+$\TOF$ using ancillae whose state is restored at the end. 
This is a folklore construction, but we present it here for self-containment.
Substitute disjunctions in $\varphi$ using De Morgan's law to obtain an expression over conjunctions and negations.
Replace each conjunction with a $\TOF$ gate and each negation with an $\XGate$ gate (implemented by a $\TOF$ gate with controls on ancillae fixed to $\ket{1}$), writing each sub-formula into a fresh ancilla. 
By uncomputing all sub-formulas in the reverse order, we obtain a reversible circuit $R_\varphi$ of size linear in $|\varphi|$ computing $\ket{y}\ket{11}\ket{b}\ket{0^s} \mapsto\ket{y}\ket{11}\ket{b\oplus\varphi(y)}\ket{0^s}$. 
Fixing $b=1$ and applying $\HGate\ket{1}=\ket{-}$ to it before applying $R_\varphi$ gives a phase kickback, contributing the phase $(-1)^{\varphi(y)}$.
Then applying a $\HGate$ gate restores it to $\ket{1}$, and since every ancilla is restored, $U_\varphi^\pm$ acts as stated on the $n$ data qubits alone.

The $\DeutschJozsa{\varphi}$ circuit computes a transformation where the amplitude on basis vectors is closely related to whether $\varphi$ is balanced.
Specifically,
\begin{align*}
    \ket{x} &\arrow{\HGate^{\otimes n}} \frac{1}{\sqrt{2^n}} \sum_{y} (-1)^{x\cdot y} \ket{y} \\
    &\arrow{U^{\pm}_\varphi} \frac{1}{\sqrt{2^n}} \sum_y (-1)^{\varphi(y)} (-1)^{x\cdot y} \ket{y} \\
    &\arrow{\HGate^{\otimes n}} \frac{1}{\sqrt{2^n}} \sum_y (-1)^{\varphi(y)} (-1)^{x\cdot y} \Big(\frac{1}{\sqrt{2^n}} \sum_z (-1)^{y\cdot z} \ket{z} \Big) \\
    &= \frac{1}{2^n} \sum_z \Big(\sum_y (-1)^{\varphi(y)}(-1)^{x\cdot y+y\cdot z} \Big) \ket{z} \\
    &= \sum_z S(x,z) \ket{z}\ ,
\end{align*}
where $S$ is
$$S(x,y):= \frac{1}{2^n}\sum_z (-1)^{\varphi(z)}(-1)^{x\cdot z+z\cdot y}\enspace.$$
Using $\DeutschJozsa{\varphi}$ as a subroutine, we construct the gadget $\DJGADGET(G,\varphi)$, shown in \cref{fig:deutsch_jozsa_gadget}.
It is parameterized by an arbitrary circuit $G$ and a Boolean formula $\varphi$.
In words, $\DJGADGET(G,\varphi)$ uses $n$ ancillae and a controlled gate $\Lambda^n G$.
In effect, $G\ket{x}$ appears in the output state if and only if $\varphi$ is not balanced.
Formally, we have the following lemma.

\begin{figure}[H]
    \centering
\begin{center}

\begin{equation*}
\Qcircuit @C=1em @R=1.6em {
   \lstick{\ket{1^n}} & {/} \qw & \gate{\DeutschJozsa{\varphi}} & \ctrl{1} & \gate{\DeutschJozsa{\varphi}} & \qw \\
   \lstick{\ket{x}}   & {/} \qw & \qw                           & \gate{G}  & \qw                                   & \qw \\
}
\end{equation*}

\end{center}
    \caption{The gadget $\DJGADGET(G,\varphi)$.}
    \label{fig:deutsch_jozsa_gadget}
\end{figure}

\smallskip
\begin{lemma}
    \label{lemma:DJ_gadget}
    Given a Boolean formula $\varphi\colon\{0,1\}^n \arrow{} \{0,1\}$ and a circuit $G$, the gadget $\DJGADGET(G,\varphi)$ computes 
    $$\DJGADGET(G,\varphi)\colon \ket{1^n} \ket{x} \mapsto \ket{\psi} \ket{x} + \beta_\varphi \ket{\psi'} (U_G\ket{x}),$$
    where $\ket{\psi'}$ is a quantum state, $\ket{\psi}$ is a vector with $||\ket{\psi}||^2=1-\beta^2_\varphi$, and $\beta_\varphi = \frac{1}{2^n} \sum_y (-1)^{\varphi(y)}$.
    Specifically, if $\varphi$ is balanced, then $\DJGADGET(G,\varphi) \colon \ket{1^n}\ket{x} \mapsto \ket{1^n}\ket{x}$.
\end{lemma}
\begin{proof}
Note that $\beta_\varphi=\frac{1}{2^n} \sum_y (-1)^{\varphi(y)}=\frac{1}{2^n}(|\{x\colon \neg\varphi(x)\}|-|\{x\colon \varphi(x)\}|)=0$ if and only if $\varphi$ is balanced.

\SubParagraph{($\varphi$ balanced).} 
The state after applying the first $\DeutschJozsa{\varphi}$ equals $\sum_z S(1^n,z)\ket{z}\ket{x}=\beta_\varphi \ket{1^n}\ket{x}+\sum_{z\neq 1^n} S(1^n,z)\ket{z}\ket{x}$, where we use that $S(1^n,1^n)=\beta_\varphi$. 
In particular, it has amplitude $\beta_\varphi=0$ on $\ket{1^n}\ket{x}$, so $\Lambda^n G$ acts as identity on the state.
Hence the circuit computes $\DeutschJozsa{\varphi} \DeutschJozsa{\varphi}$. 
Since $\DeutschJozsa{\varphi} = \HGate^{\otimes n}U_\varphi^\pm \HGate^{\otimes n}$ with $\HGate^2=\IGate$ and $(U_\varphi^\pm)^2=\IGate$, the circuit $\DeutschJozsa{\varphi}$ is an involution, so $\DeutschJozsa{\varphi} \DeutschJozsa{\varphi} = \IGate$.

\SubParagraph{($\varphi$ not balanced).} 
In particular, $\beta_\varphi \neq 0$, so $\DJGADGET(G,\varphi)$ computes
\begin{align*}
    \ket{1^n}\ket{x} &\arrow{\DeutschJozsa{\varphi}} \sum_z S(1^n,z) \ket{z}\ket{x} \\
    &\arrow{\Lambda^n G} \sum_{z\neq 1^n} S(1^n,z) \ket{z}\ket{x} + S(1^n,1^n) \ket{1^n}(U_G\ket{x}) \\
    &\arrow{\DeutschJozsa{\varphi}} \sum_{z\neq 1^n} S(1^n,z) \sum_w S(z,w) \ket{w} \ket{x} + \beta_\varphi \sum_w S(1^n,w) \ket{w} (U_G\ket{x}) \\
    & =\ket{\psi} \ket{x} + \beta_\varphi \ket{\psi'} (U_G\ket{x})\ ,
\end{align*}
where
$\ket{\psi}:= \sum_{z\neq 1^n} S(1^n,z) \sum_w S(z,w) \ket{w}$ and $\ket{\psi'} := \sum_w S(1^n,w) \ket{w}$.
Since $\DeutschJozsa{\varphi}$ is unitary, $\ket{\psi'} = \DeutschJozsa{\varphi}\ket{1^n}$ is a quantum state, and $\ket{\psi}$ is the image under $\DeutschJozsa{\varphi}$ of $\DeutschJozsa{\varphi}\ket{1^n}$ with its component on $\ket{1^n}$ deleted, so $||\ket{\psi}||^2=1-|S(1^n,1^n)|^2=1-\beta_\varphi^2$.
\end{proof}

\section{The Hardness of Exact Circuit Optimization}\label{sec:hardness}

We now present our key technical lemma, which establishes the complexity of $\ISIDENTITYPROMISE{\gateset}$.
After this, we give the main theorem, establishing the hardness of exact quantum circuit optimization.

\smallskip
\begin{lemma}
    \label{lemma:promise_problem_hardness}
    For any gate-set $\gateset$ that has ancilla-restoring exact implementations of $\HGate$ and $\TOF$, $\ISIDENTITYPROMISE{\gateset}$ is $\CO{\NQP}$-hard.
\end{lemma}
\begin{proof}
    We establish a Karp reduction from $\ISBALANCED$ to $\ISIDENTITYPROMISE{\text{$\gateset$}}$.
    
    Given a Boolean formula $\varphi$ over $n$ bits, we construct the circuit $C$ shown in~\cref{fig:G_gadget}.
    The circuit $C$ is composed of a $\DJGADGET(\TOF,\varphi)$ gadget and four $\HGate$ gates and hence not necessarily over $\gateset$.

\begin{figure}[H]
    \centering
\begin{center}

\begin{equation*}
\Qcircuit @C=1em @R=1.6em {
   \lstick{\ket{1^{n}}} & {/} \qw & \gate{\DeutschJozsa{\varphi}} & \ctrl{1} & \gate{\DeutschJozsa{\varphi}} & \qw \\
   \lstick{\ket{0}}   &     \qw & \gate{\HGate}                 & \ctrl{1} & \gate{\HGate}                 & \qw \\
   \lstick{\ket{0}}   &     \qw & \gate{\HGate}                 & \ctrl{1} & \gate{\HGate}                 & \qw \\
   \lstick{\ket{0}}   &     \qw & \qw                           & \targ{}  & \qw                           & \qw \\
}
\end{equation*}

\end{center}
    \caption{A circuit describing $C$ given $\varphi$.}
    \label{fig:G_gadget}
\end{figure}

    Towards a circuit $C''$ over $\gateset$ implementing $C$, we construct a $\HGate$+$\TOF$ circuit $C'$ implementing $C$, as follows: Construct $\DeutschJozsa{\varphi}$ over $\HGate$+$\TOF$ with clean ancillae as in~\cref{sec:DJ_gadget}, and the multi-controlled $\TOF$ gate from ordinary $\TOF$ gates and $O(n)$ clean ancillae by the standard construction of~\cite{Barenco1995ElementaryGatesForQuantumComputation}.
    To obtain $C''$, substitute all $\HGate$ resp. $\TOF$ in $C'$ by their ancilla-restoring implementations over $\gateset$.
    Then $C''$ is an ancilla-restoring circuit over $\gateset$.
    Let $y=y_1 y_2 \dots y_k$ for $y_j\in \{0,1\}$ be the initial values of the ancillae of $C''$.
    Then for all $\ket{\psi}$ ranging over the data qubits, $U_{C''}\ket{\psi}\ket{y}=(U_C\ket{\psi})\ket{y}$.
    We pick $C''$ together with $z:=1^n000y$ as our yes/no instance of $\ISIDENTITYPROMISE{\text{$\gateset$}}$.

    As $U_{C''}\ket{z}=(U_{C}\ket{1^n}\ket{000})\ket{y}$, it suffices to argue about $(U_{C}\ket{1^n}\ket{000})\ket{y}$.
    We argue that if $\varphi$ is balanced, then $(U_C\ket{1^{n}}\ket{000})\ket{y}=\ket{1^{n}}\ket{000}\ket{y}$, and if not, then $(U_C\ket{1^{n}}\ket{000})\ket{y}$ is a superposition state exhibiting correlation and a biased outcome probability.
    Since deciding whether $\varphi$ is balanced is hard for $\CEQP$~\cite{wagner1986complexity} and $\CEQP=\CO{\NQP}$~\cite{Yamakami1999nqp}, this will conclude the proof.
    We now argue about the correctness of the construction. 
    
    (\SubParagraph{$\varphi$ balanced $\Rightarrow$ $(U_C\ket{1^{n}}\ket{000})\ket{y}=\ket{1^{n}}\ket{000}\ket{y}$}).
    We apply \cref{lemma:DJ_gadget} on $\DJGADGET(\TOF,\varphi)$ to see that $\DJGADGET(\TOF, \varphi) \colon \ket{1^{n}}\ket{x} \mapsto \ket{1^{n}}\ket{x}$ while restoring all ancillae, so the $\HGate$ gates cancel out. 
    Hence, $(U_C\ket{1^{n}}\ket{000})\ket{y}=\ket{1^{n}}\ket{000}\ket{y}$.

    (\SubParagraph{$\varphi$ not balanced $\Rightarrow$ $(U_C\ket{1^{n}}\ket{000})\ket{y}$ exhibits superposition, correlation, and  biased outcome probability}).
    By~\cref{lemma:DJ_gadget}, as $\varphi$ is not balanced, there exist a constant $\beta_\varphi$ and vectors $\ket{\psi}$ and $\ket{\psi'}$ such that $\beta_\varphi\neq 0$, and $C$ on $\ket{1^{n}}\ket{000}$ computes the following. 
        \begin{align*}
        \ket{1^{n}}\ket{000} 
            &\arrow{\HGate^{\otimes 2}} \frac{1}{2} \ket{1^n} (\ket{00}+\ket{01}+\ket{10}+\ket{11})\ket{0} \\
            &\arrow{\DJGADGET(\TOF,\varphi)} \frac{1}{2}\ket{\psi} (\ket{00}+\ket{01}+\ket{10}+\ket{11})\ket{0}  \\
            & \quad + \frac{\beta_{\varphi}}{2} \ket{\psi'} (\ket{000}+\ket{010}+\ket{100}+\ket{111}) \\
            &\arrow{\HGate^{\otimes 2}} \ket{\psi} \ket{000} + \frac{\beta_{\varphi}}{4}\ket{\psi'}(3\ket{000}+\ket{010}+\ket{100} \\
            & \quad -\ket{110}+\ket{001}-\ket{011}-\ket{101}+\ket{111}) \ .
    \end{align*}

    (Superposition). 
    The state has (among others) non-zero elements $\ket{\psi'}\ket{010}\ket{y}$, $\ket{\psi'}\ket{100}\ket{y}$, and $\ket{\psi'}\ket{111}\ket{y}$, none of which can cancel out with another, so at least two basis states have non-zero amplitude, hence $(U_C\ket{1^{n}}\ket{000})\ket{y}$ is a superposition state.

    (Correlation and biased outcome probability). 
    Consider the last three qubits and call them $a$, $b$, $c$. 
    Every outcome with $abc\neq 000$ has probability $\beta_\varphi^2/16$, and the outcome $000$ has probability $1-7\beta_\varphi^2/16$. 
    Hence
    $P(c=1|ab=01) = 1/2$,
    whereas
    $$P(c=1|ab=00) = \frac{\beta_\varphi^2/16}{(1-7\beta_\varphi^2/16) + \beta_\varphi^2/16} = \frac{\beta_\varphi^2}{16-6\beta_\varphi^2} \in \left(0, \tfrac{1}{10}\right] \ ,$$
    using $0 < \beta_\varphi^2 \leq 1$. 
    Thus $P(c=1|ab=01)\neq P(c=1|ab=00)$, hence $(U_C\ket{1^{n}}\ket{000})\ket{y}$ exhibits correlation.
    Further, $P(c=1|ab=00)\notin\{0,1/2,1\}$, hence $(U_C\ket{1^{n}}\ket{000})\ket{y}$ has a biased outcome probability.

    In particular, the constructed instance satisfies the promise, and it is a yes-instance if and only if $\varphi$ is balanced.
\end{proof}

\Paragraph{Consequences on optimization.}
We now give the main theorem.
For gate-sets $\gateset$ and $\gatesetB$, and objective function $f$, we define the problem $\OPTIMIZATIONPROBLEM{\gateset, \gatesetB, f}$. 

\titlebox{$\OPTIMIZATIONPROBLEM{\gateset, \gatesetB, f}$}{
\emph{Input:} An $n$-qubit circuit $C_\gateset$ over gate-set $\gateset$, bit-string $x$ with $|x|\leq n$, integer $k$. \\
\emph{Question:} Does there exist a circuit $C_\gatesetB$ over $\gatesetB$ such that $f(C_\gatesetB)\leq k$, and for all $(n-|x|)$-qubit states $\ket{\psi}$, it holds that $U_{C_\gateset}\ket{x}\ket{\psi}=U_{C_\gatesetB}\ket{x}\ket{\psi}$?
}

Here $C_\gateset$ is a specification circuit over gate-set $\gateset$, and we ask for an implementation $C_\gatesetB$ over gate-set $\gatesetB$ within budget $k$ according to a cost function $f$ that assigns a real number to each circuit.  
We require that the implementation $C_\gatesetB$ is equal to $C_\gateset$ on the subspace spanned by states of the form $\ket{x}\ket{\psi}$ for all $\ket{\psi}$.

\smallskip
\begin{theorem}
\label{theorem:optimization_problem_hardness}
$\OPTIMIZATIONPROBLEM{\gateset, \gatesetB, f}$ where $\gateset$ is any gate-set with exact, ancilla-restoring implementations of $\HGate$ and $\TOF$, and $\gatesetB$ is any gate-set, and $f$ measures the count or depth of
(i)~all gates, or
(ii)~non-Clifford gates, or
(iii)~superposition gates, or
(iv)~entanglement gates,
is $\CO{\NQP}$-hard.
\end{theorem}

\begin{proof}
    We show that $\OPTIMIZATIONPROBLEM{\gateset, \gatesetB, f}$ can be used to decide $\ISIDENTITYPROMISE{\text{$\HGate$+$\TOF$}}$, and hence inherits the hardness given by~\cref{lemma:promise_problem_hardness}.

    Let $(C,x)$ be a yes/no instance of $\ISIDENTITYPROMISE{\text{$\HGate$+$\TOF$}}$. 
    Note that $C$ is a $\HGate$+$\TOF$ circuit. 
    We construct a corresponding yes/no instance of $\OPTIMIZATIONPROBLEM{\gateset, \gatesetB, f}$ as follows.
    Let $C_\gateset$ be a circuit over $\gateset$, constructed from $C$ by replacing each $\HGate$ resp. $\TOF$ gate with the corresponding gadget over $\gateset$ that implements it.
    This can be done in polynomial time.
    Then let $(C_\gateset,x,k=0)$ be the yes/no instance of $\OPTIMIZATIONPROBLEM{\gateset, \gatesetB, f}$. 

    It remains to show that $(C,x)$ is a yes-instance of $\ISIDENTITYPROMISE{\text{$\HGate$+$\TOF$}}$ if and only if $(C_\gateset,x,k=0)$ is a yes-instance of $\OPTIMIZATIONPROBLEM{\gateset, \gatesetB, f}$.

    Since by construction, $U_{C_\gateset}\ket{x}=U_{C}\ket{x}$, then $(C_\gateset,x,k=0)$ is a yes-instance if and only if there exists a circuit $C_\gatesetB$ over $\gatesetB$ with
    \begin{align}
        U_{C}\ket{x} = U_{C_\gatesetB} \ket{x} \quad \text{and} \quad f(C_\gatesetB)=0 \ . \label{eq:proof_cond}
    \end{align}

    Suppose $(C,x)$ is a yes-instance, i.e. $U_C\ket{x} = \ket{x}$. 
    Take $C_\gatesetB$ to be the empty circuit, so that $U_{C_\gatesetB}\ket{x} = \ket{x} = U_C \ket{x}$. 
    The empty circuit contains no gates at all, so its count and its depth are $0$ for each of the resource types (i)–(iv). 
    Hence~\cref{eq:proof_cond} holds and $(C_\gateset,x,k=0)$ is a yes-instance.
    Note further that the empty circuit can be implemented over any $\gatesetB$.

    Suppose instead that $(C,x)$ is a no-instance. 
    By the promise, $U_C\ket{x}$ is a superposition state that exhibits correlation and has a biased outcome probability. 
    Suppose there exists a circuit $C_\gatesetB$ over $\gatesetB$ with $U_{C_\gatesetB}\ket{x} = U_C\ket{x}$; if not, then $(C_\gateset,x,k=0)$ is already a no-instance. 
    By~\cref{lemma:state_obtained_by_circuit_gives_property_of_circuit}, $C_\gatesetB$ contains at least one superposition gate, at least one entanglement gate and at least one non-Clifford gate.
    Hence, for every resource type (i)–(iv), the circuit $C_\gatesetB$ contains at least one gate of that type.
    A single such gate forms a chain of length one, so both the count and the depth of that resource are at least $1$, giving $f(C_\gatesetB)\geq 1$.
    As $C_\gatesetB$ was arbitrary, no circuit satisfies~\cref{eq:proof_cond}, and $(C_\gateset,x,k=0)$ is a no-instance.

    From~\cref{lemma:promise_problem_hardness}, $\CO{\NQP}$-hardness follows.
\end{proof}

Note that $\gatesetB$ does not need to implement any of the gates of $\gateset$, as the proof only relies on the fact that any $\gatesetB$ implements the identity.
Further, note that the proof goes through when $C_\gatesetB$ may use clean or borrowed ancillae, since such ancillae must be restored and tensoring with a basis state preserves superposition, correlation, and biased outcome probability.
The conditions of~\cref{theorem:optimization_problem_hardness} are in particular satisfied by the Clifford+$\TGate$ gate-set.

\smallskip
\begin{corollary}
    \label{corollary:1}
    Let $\gateset$ be any gate-set with exact, ancilla-restoring implementations of $\HGate$ and $\TOF$, and let $\gatesetB$ be any gate-set.
    With respect to equivalence on a subspace, each of the following is $\CO{\NQP}$-hard: minimizing (a)~$\TGate$-count or $\TGate$-depth of Clifford+$\TGate$ implementations; (b)~$\HGate$-count or $\HGate$-depth over Clifford+$\TGate$ or over $\HGate$+$\TOF$; (c)~$\CXGate$-count or $\CXGate$-depth over Clifford+$\TGate$; (d)~$\TOF$-count or $\TOF$-depth over $\HGate$+$\TOF$; (e)~total gate count or circuit depth over $\gatesetB$.
\end{corollary}

\begin{proof}
    Each case follows from~\cref{theorem:optimization_problem_hardness}, since the gate being counted is the only gate of the relevant resource type in the gate-set at hand. 
    $\TGate$ is the only non-Clifford gate of Clifford+$\TGate$, giving (a) from case (ii). 
    $\HGate$ is the only superposition gate of Clifford+$\TGate$ and of $\HGate$+$\TOF$, giving (b) from case (iii). 
    $\CXGate$ is the only entangling gate of Clifford+$\TGate$ and $\TOF$ the only entangling gate of $\HGate$+$\TOF$, giving (c) and (d) from case (iv); (d) also follows from case (ii), as $\TOF$ is the only non-Clifford gate of $\HGate$+$\TOF$. 
    Case (e) is case (i). 
\end{proof}

These are the resource measures that dominate practice. 
In fault-tolerant architectures, non-Clifford gates are produced by magic state distillation, whose cost typically dominates the space-time volume of a computation, which makes $\TGate$-count and $\TGate$-depth the standard metrics in that setting~\cite{Litinski_2019}; 
they are also the metrics for which the $\NP$-hardness bounds of~\cite{Wetering2023OptimisingQuantumCircuitsHard} were obtained. 
Hadamard gates are the principal obstruction to phase-polynomial based optimization, as they break a circuit into blocks that must be handled separately, so $\HGate$-count is a natural secondary target~\cite{Amy_2014}. 
On near-term hardware, two-qubit entangling gates carry error rates one to two orders of magnitude above typical single-qubit gates~\cite{IBMQuantum2025_QPUInformation,IonQ2025_NativeGatesGuide}, motivating $\CXGate$-count minimization.
\cref{corollary:1} shows that exact minimization of each of these metrics is $\CO{\NQP}$-hard, and hence outside $\PH$ unless $\PH$ collapses.

\Paragraph{Relation to existing work.}
It is known that $\CO{\NQP}$-hard problems lie outside $\PH$ unless $\PH$ collapses, since $\CO{\NQP}=\CEQP$~\cite{Yamakami1999nqp} and $\CEQP$ has this property~\cite{TodaOgiwara1992}. 
Overall, \cref{theorem:optimization_problem_hardness} complements the recent $\NP$-hardness lower bounds of~\cite{Wetering2023OptimisingQuantumCircuitsHard}, and tightens the gap to the corresponding $\NP^{\NQP}$ upper bound proven in that work.
In addition, \cref{theorem:optimization_problem_hardness} generalizes the $\CO{\NQP}$-hardness bound of~\cite{Tanaka2010ExactNonIdentityCheckNQPComplete} to circuits over fixed gate families and when minimization is with respect to specific gate fragments, each expressing a different type of quantum resource, as opposed to all gates.

\section{Conclusion}

We have shown that minimizing for depth or count over superposition, entanglement, non-Clifford, or total gate resources are all hard problems for $\CO{\NQP}$, and hence outside of $\PH$ unless $\PH$ collapses.
Our result applies to any gate-set that implements $\HGate$+$\TOF$ exactly, including standard gate-sets such as Clifford+$\TGate$.
Finally, the reduction behind \cref{theorem:optimization_problem_hardness} is based on a small and simple gadget.
This indicates that strong hardness can arise out of standard quantum operations, and hints that it may be prevalent in practice.

\section*{Acknowledgements}

This work was partially supported by a research grant (VIL42117) from VILLUM FONDEN.

\bibliography{references}

@article{Aaronson04sim_of_stabilizer_circuits,
   title={{Improved simulation of stabilizer circuits}},
   volume={70},
   ISSN={1094-1622},
   DOI={10.1103/physreva.70.052328},
   number={5},
   journal={Physical Review A},
   publisher={American Physical Society (APS)},
   author={Aaronson, Scott and Gottesman, Daniel},
   year={2004},
   month=nov }

@article{TodaOgiwara1992,
  author       = {Seinosuke Toda and Mitsunori Ogiwara},
  title        = {{{C}ounting {C}lasses are at {L}east as {H}ard as the {P}olynomial‐{T}ime {H}ierarchy}},
  journal      = {SIAM Journal on Computing},
  year         = {1992},
  volume       = {21},
  number       = {2},
  pages        = {316--328},
  doi          = {10.1137/0221023},
}

@inproceedings{Schneider2023,
author = {Schneider, Sarah and Burgholzer, Lukas and Wille, Robert},
title = {{A SAT Encoding for Optimal Clifford Circuit Synthesis}},
year = {2023},
isbn = {9781450397834},
publisher = {Association for Computing Machinery},
address = {New York, NY, USA},
doi = {10.1145/3566097.3567929},
booktitle = {Proceedings of the 28th Asia and South Pacific Design Automation Conference},
pages = {190–195},
numpages = {6},
location = {Tokyo, Japan},
series = {ASPDAC '23}
}

@article{Litinski_2019,
   title={{Magic State Distillation: Not as Costly as You Think}},
   volume={3},
   ISSN={2521-327X},
   DOI={10.22331/q-2019-12-02-205},
   journal={Quantum},
   publisher={Verein zur Forderung des Open Access Publizierens in den Quantenwissenschaften},
   author={Litinski, Daniel},
   year={2019},
   month=Dec, pages={205} }

@InProceedings{shaik2025cnotoptimalcliffordsynthesissat,
      title={{CNOT-Optimal Clifford Synthesis as SAT}}, 
      author={Irfansha Shaik and Jaco van de Pol},
  booktitle =	{28th International Conference on Theory and Applications of Satisfiability Testing (SAT 2025)},
  pages =	{28:1--28:21},
  series =	{Leibniz International Proceedings in Informatics (LIPIcs)},
  ISBN =	{978-3-95977-381-2},
  ISSN =	{1868-8969},
  year =	{2025},
  volume =	{341},
  editor =	{Berg, Jeremias and Nordstr\"{o}m, Jakob},
  publisher =	{Schloss Dagstuhl -- Leibniz-Zentrum f{\"u}r Informatik},
  address =	{Dagstuhl, Germany},
  URN =		{urn:nbn:de:0030-drops-237621},
  doi =		{10.4230/LIPIcs.SAT.2025.28}
}

@article{Bravyi2022_6_qubit_optimal_Clifford_circuits,
   title={{6-qubit optimal {C}lifford circuits}},
   volume={8},
   ISSN={2056-6387},
   DOI={10.1038/s41534-022-00583-7},
   number={1},
   journal={npj Quantum Information},
   publisher={Springer Science and Business Media LLC},
   author={Bravyi, Sergey and Latone, Joseph A. and Maslov, Dmitri},
   year={2022},
   month=jul }

@article{Dawson2005solovay_kitaev_thrm,
    author = {Dawson, Christopher M. and Nielsen, Michael A.},
    title = {{The Solovay-Kitaev algorithm}},
    year = {2006},
    issue_date = {January 2006},
    publisher = {Rinton Press, Incorporated},
    address = {Paramus, NJ},
    volume = {6},
    number = {1},
    issn = {1533-7146},
    journal = {Quantum Info. Comput.},
    month = jan,
    pages = {81–95},
    numpages = {15},
    url={https://dl.acm.org/doi/abs/10.5555/2011679.2011685}, 
}

@book{Nielsen_Chuang2010,
  Author = {Michael A. Nielsen and Isaac L. Chuang},
  Title = {{Quantum Computation and Quantum Information: 10th Anniversary Edition}},
  Publisher = {Cambridge University Press},
  Year = {2011},
  ISBN = {9781107002173}
}

@book{Qiu2025,
  author = {Daowen Qiu},
  title = {{Theoretical Foundations of Quantum Computing}},
  publisher = {Morgan Kaufmann},
  address = {Cambridge, MA},
  year = {2025}
}

@article{DeutschJozsa1992RapidSolutionProblemsQuantumComputation,
  title={{Rapid solution of problems by quantum computation}},
  author={Deutsch, David and Jozsa, Richard},
  journal={Proceedings of the Royal Society of London. Series A: Mathematical and Physical Sciences},
  volume={439},
  number={1907},
  doi={https://doi.org/10.1098/rspa.1992.0167},
  pages={553--558},
  year={1992},
  publisher={The Royal Society London}
}

@article{Barenco1995ElementaryGatesForQuantumComputation,
   title={{Elementary gates for quantum computation}},
   volume={52},
   ISSN={1094-1622},
   DOI={10.1103/physreva.52.3457},
   number={5},
   journal={Physical Review A},
   publisher={American Physical Society (APS)},
   author={Barenco, Adriano and Bennett, Charles H. and Cleve, Richard and DiVincenzo, David P. and Margolus, Norman and Shor, Peter and Sleator, Tycho and Smolin, John A. and Weinfurter, Harald},
   year={1995},
   month=nov, pages={3457–3467} }

@article{Amy_2014,
   title={{Polynomial-Time T-Depth Optimization of Clifford+T Circuits Via Matroid Partitioning}},
   volume={33},
   ISSN={1937-4151},
   DOI={10.1109/tcad.2014.2341953},
   number={10},
   journal={IEEE Transactions on Computer-Aided Design of Integrated Circuits and Systems},
   publisher={Institute of Electrical and Electronics Engineers (IEEE)},
   author={Amy, Matthew and Maslov, Dmitri and Mosca, Michele},
   year={2014},
   month=Oct, pages={1476–1489} }

@article{Muroya2025HardwareOptimal,
  author    = {Muroya, S. and Chatterjee, K. and Henzinger, T. A.},
  title     = {{Hardware-optimal quantum algorithms}},
  journal   = {Proceedings of the National Academy of Sciences of the United States of America},
  year      = {2025},
  volume    = {122},
  number    = {12},
  pages     = {e2419273122},
  doi       = {10.1073/pnas.2419273122},
  pmid      = {40106357},
  pmcid     = {PMC11962418}
}

@article{Karuppasamy2025QuantumCircuitOptimization,
  author       = {Krishnageetha Karuppasamy and Varun Puram and Stevens Johnson and Johnson P. Thomas},
  title        = {{A Comprehensive Review of Quantum Circuit Optimization: Current Trends and Future Directions}},
  journal      = {Quantum Reports},
  year         = {2025},
  volume       = {7},
  number       = {1},
  doi          = {10.3390/quantum7010002},
  issn         = {2624-960X}
}

@article{Wetering2023OptimisingQuantumCircuitsHard,
  title={{Optimising quantum circuits is generally hard}},
  author={van de Wetering, John and Amy, Matt},
  doi={https://doi.org/10.48550/arXiv.2310.05958},
  journal={arXiv preprint arXiv:2310.05958},
  year={2023}
}

@article{dehaene2003clifford,
  title={{Clifford group, stabilizer states, and linear and quadratic operations over {GF}(2)}},
  author={Dehaene, Jeroen and De Moor, Bart},
  journal={Physical Review A},
  volume={68},
  number={4},
  pages={042318},
  year={2003},
  publisher={APS},  
  DOI={10.1103/physreva.68.042318},
}

@article{Tanaka2010ExactNonIdentityCheckNQPComplete,
  title={{Exact non-identity check is {NQP}-complete}},
  author={Tanaka, Yu},
  journal={International Journal of Quantum Information},
  doi={https://doi.org/10.1142/S0219749910006599},
  volume={8},
  number={05},
  pages={807--819},
  year={2010},
  publisher={World Scientific}
}

@misc{fenner1998determiningacceptancepossibilityquantum,
      title={{Determining Acceptance Possibility for a Quantum Computation is Hard for the Polynomial Hierarchy}}, 
      author={Stephen Fenner and Frederic Green and Steven Homer and Randall Pruim},
      year={1998},
      eprint={quant-ph/9812056},
      archivePrefix={arXiv},
      primaryClass={quant-ph},
}

@article{Yamakami1999nqp,
  title={{{NQP} = co-{C$_=$P}}},
  author={Yamakami, Tomoyuki and Yao, Andrew C},
  journal={Information Processing Letters},
  doi = {https://doi.org/10.1016/S0020-0190(99)00084-8},
  volume={71},
  number={2},
  pages={63--69},
  year={1999},
  publisher={Elsevier}
}

@article{aharonov2003simple,
  title={{A simple proof that {T}offoli and {H}adamard are quantum universal}},
  author={Aharonov, Dorit},
  journal={arXiv preprint quant-ph/0301040},
  year={2003},
  doi={https://doi.org/10.48550/arXiv.quant-ph/0301040}
}

@article{adleman1997quantum,
  author    = {Leonard M. Adleman and Jonathan DeMarrais and Ming-Deh A. Huang},
  title     = {{Quantum {C}omputability}},
  journal   = {SIAM Journal on Computing},
  volume    = {26},
  number    = {5},
  pages     = {1524--1540},
  year      = {1997},
  doi       = {10.1137/S0097539795287997}
}

@article{wagner1986complexity,
  title={{The complexity of combinatorial problems with succinct input representation}},
  author={Wagner, Klaus W},
  journal={Acta Informatica},
  doi={https://doi.org/10.1007/BF00289117},
  volume={23},
  number={3},
  pages={325--356},
  year={1986},
  publisher={Springer}
}

@misc{IBMQuantum2025_QPUInformation,
  author       = {{IBM Corporation}},
  title        = {{QPU information}},
  howpublished = {\url{https://quantum.cloud.ibm.com/docs/en/guides/qpu-information}},
  note         = {Accessed: 2025-10-13},
  year         = {2025}
}

@misc{IonQ2025_NativeGatesGuide,
  author       = {{IonQ}},
  title        = {{Getting started with IonQ’s hardware-native gateset}},
  howpublished = {\url{https://docs.ionq.com/guides/getting-started-with-native-gates}},
  note         = {Accessed: 2025-10-13},
  year         = {2025}
}

\end{document}